\documentstyle[aas2pp4]{article}

\received{}
\accepted{}
\journalid{}{}
\articleid{}{}

\slugcomment{}

\lefthead{Yu et al.}
\righthead{Shock-Excited Flows in OMC-2 and OMC-3}

\begin{document}

\title{Shock-Excited H$_2$ Flows in OMC-2 and OMC-3}

\author{Ka Chun Yu\altaffilmark{1,4,5}, J. Bally\altaffilmark{2,5}, and
    David Devine\altaffilmark{3,4}}
\affil{Department of Astrophysics, Planetary, and Atmospheric Sciences
 \\ and Center for Astrophysics and Space Astronomy}
\authoraddr{University of Colorado, Boulder, CO 80309-089}

\altaffiltext{1}{kachun@casa.colorado.edu}
\altaffiltext{2}{bally@nebula.colorado.edu}
\altaffiltext{3}{devine@starburst.colorado.edu}
\altaffiltext{4}{Visiting Astronomer, Kitt Peak National Observatory. 
KPNO is operated by AURA, Inc.\ under contract to the National Science
Foundation.} 
\altaffiltext{5}{Visiting Astronomer, Cerro-Tololo Inter-American Observatory. 
CTIO is operated by AURA, Inc.\ under contract to the National Science
Foundation.} 

\begin{abstract} 
We report the discovery of nearly a dozen collimated outflows from
young stellar objects embedded in the molecular filament that extends
north of the Orion Nebula towards the H~II region NGC~1977.  The
large number of nearly co-eval outflows and embedded class-0 young
stellar objects indicates that the OMC-2/3 region is one of the most
active sites of on-going low to intermediate mass star formation
known.  These outflows were identified in the 2.12~$\mu$m $v=1$--$0 ~S(1)$
H$_2$ line during a survey of a 6\arcmin\ $\times$ 16\arcmin\ region
containing the OMC-2 and OMC-3 cloud cores and over a dozen recently
discovered class-0 protostars.  We also observe filamentary emission
that is likely to trace possible fluorescent H$_2$ in
photo-dissociation regions associated with M~43 and NGC~1977.
Neither the suspected outflows nor the fluorescent emission are seen
at the continuum wavelength of 2.14~$\mu$m which confirms their
emission line nature.  Several of the new H$_2$ flows are associated
with recently discovered bipolar molecular outflows.  However, the
most prominent bipolar CO outflow from the region (the MMS~8 flow)
has no clear H$_2$ counterpart.  Several H$_2$ flows consist of
chains of knots and compact bow shocks that likely trace highly
collimated protostellar jets.  Our discovery of more than 80
individual H$_2$ emitting shocks demonstrate that outflows from young
stars are churning this molecular cloud.  \end{abstract}


\section{Introduction}

Large format arrays operating in the near-infrared (NIR) at a
wavelength of 2~$\mu$m provide an unprecedented opportunity to probe
the interior of molecular clouds where the youngest young stellar
objects (class-0 YSOs) are found.  In these highly obscured
environments the powerful outflows produced by YSOs can be identified
by the 2.12~$\mu$m emission from the $v=1$--$0 ~S(1)$ line of H$_2$
excited in shocks (e.g., Bally et al.\ 1993; McCaughrean, Rayner, \&
Zinnecker 1994; Hodapp \& Ladd 1995; Davis \& Eisl\"offel 1995;
Zinnecker, McCaughrean, \& Rayner 1997).  Imaging at 2.12~$\mu$m can
lead to the discovery of outflows that may otherwise not be found via
the identification of optical Herbig-Haro (HH) objects due to high
obscuration, or by means of millimeter-wavelength maps of molecular
outflows due to poor angular resolution.

OMC-2 and OMC-3 are two major star forming cloud cores north of
OMC-1, located at a distance of about 480~pc, which form a molecular
ridge bounded by the Orion Nebula to the south, and NGC~1977 to the
north.  OMC-2 and some of its infrared sources were identified by
Gatley (1974) and subsequently investigated in the near- and
far-infrared (FIR) by Cohen \& Frogel (1977), Thronson et al.\
(1978), Thronson \& Thompson (1982), and Pendleton et al.\ (1986).
NIR surveys in $J$, $H$, and $K$ bands by Johnson et al.\ (1990) and
Jones et al.\ (1994) demonstrated that the majority of stars along
this line of sight are young and likely members of the Orion OB
association.  The cloud was mapped in $^{13}$CO by Bally et al.\
(1987), in CS by Tatematsu et al.\ (1993), in C$^{18}$O by Dutrey
et al.\ (1993), and in NH$_3$ by Cesaroni \& Wilson (1994).  Images
of continuum emission by dust at 1.3~mm of OMC-2 (Mezger, Wink, \&
Zylka 1990) and both OMC-2 and OMC-3 with 19\arcsec ~resolution
(Chini et al.\ 1997) show a remarkable north-south chain of over a
dozen class-0 protostars embedded in a thin dust filament extending
from OMC-2 to OMC-3.  This unusual concentration of embedded
proto-stars motivated our NIR imaging program.

\section{Observations}

We observed the OMC-2 and OMC-3 region with the Infrared Imager
(IRIM) at the KPNO 2.1~m on 30 November and 2 December 1996.  IRIM is
a $256\times256$ HgCdTe NICMOS~3 array (Cooper et al.\ 1993; Rieke et
al.\ 1993), in an up-looking cryostat cooled by liquid nitrogen in a
dual reservoir system that separately cools the optics and the
detector.  The detector has an RMS readout noise of $\sim 37$~e$^-$
pix$^{-1}$.  The f$/15$ configuration on the 2.1~m gives a scale of
1.09 arcsec/pixel, or a field of view of approximately $280 \times
280$~arcsec$^2$.  Additional data were obtained on the CTIO 1.5~m on
February 22--23, 1997, using the $256\times256$ CTIO Infrared Imager
(CIRIM).  This system operates at f$/7.5$ providing a pixel scale of
1.14 arcsec/pixel but is functionally identical to the IRIM system at
KPNO.  The 1\% narrow-band filters are matched in bandpass and wedged
to decrease fringing from night sky lines.

We obtained a trio of dithered images in the 2.12~$\mu$m $v=1$--$0 ~S(1)$
H$_2$ filter at each pointing.  A trio of dithered sky frames
were taken for every two on-source pointings.  These were median
combined to create a `sky' frame which was subtracted from the
on-source images.  A normalized flat field was formed from dark
subtracted sky frames and divided into the images which were median
combined into the final mosaic.  A 2.14~$\mu$m continuum image was
formed in a similar manner except that on-source images were median
combined to create sky frames for the first night of observations.
This lead to an inferior continuum image, but this result did not
hamper identification of H$_2$ features.  All images were exposed for
300~seconds.  We used the UKIRT provisional faint standard (Casali \&
Hawarden 1992) SA~96-83 for flux calibration.

Astrometry for the final mosaiced field was obtained by using the
results from Ali \& Depoy's (1995) 2.2~$\mu$m $K$-band survey of the
Orion~A molecular cloud region, which covered a $39\arcmin \times
39\arcmin $ area centered around the Trapezium at a resolution of
1.5$\arcsec $.  A plate solution was found using the PLTSOL command
in the STSDAS package in IRAF.  The Ali \& Depoy survey did not
extend north of $\delta = -5^{\circ}04\arcmin $, so the northern
quarter of our field did not contain stars with known coordinates.
We compared the positions of stars (determined with IRAF's RIMCURSOR)
used in our plate solution with coordinates from Ali \& Depoy to
estimate the plate solution error, which is $\lesssim 0.5\arcsec $ in
the southern half of our field, and $\sim 1.0\arcsec $ for the
northernmost Ali \& Depoy stars.

\section{Results}

Figures~1, 2, 3, and 4 (Plates 000 through 000) respectively show the
2.14~$\mu$m continuum image, 2.12~$\mu$m line image, a difference
image obtained by subtracting the 2.14~$\mu$m image from the
2.12~$\mu$m image, and enlarged sections of the 2.12~$\mu$m image.
The locations of known {\it IRAS}, far-infrared (FIRS), 1.3 mm
continuum (MMS), and other infrared sources identified in OMC-2 (IRS)
are also shown.  Table~1 lists the locations, ordered by R.A.\ and
Dec., of the H$_2$ knots and shocks along with possible associations
with suspected outflows.  Figure~4 shows lines indicating the
locations of the suspected collimated outflows.  Suspected outflows
and their driving YSOs are clustered at three locations.

\subsection{Flows near OMC-3}  The part of our field north of $\delta
= -5^{\circ} 04\arcmin $ contains three or four possible H$_2$
outflows as well as a set of streamers that form a fan-shaped pattern
at the northern edge of our field.  Two prominent H$_2$ outflows are
oriented roughly east-west.  Flow $B$ consists of knots 1, 2, 3, and
7 and is likely powered by a member of the binary MMS~2/3.  This
chain of H$_2$ knots coincides with Herbig-Haro object HH~331
(Reipurth 1997; hereafter R97).  Flow $C$ consists of a bright H$_2$
jet (knots 13, 17, and possibly 28) and may be powered by MMS~5.
The Herbig-Haro object HH~293 (Reipurth, Bally, \& Devine 1997;
hereafter RBD97) coincides with the H$_2$ jet.  Several faint knots
of H$_2$ (Nos.\ 32, 30, and possibly 28) lie on a north-south axis
(flow $A$) centered on MMS~6, the brightest 1.3~mm source in OMC-3.
Finally, there is evidence of two flows running parallel to axis
$D$.  A faint knotty streamer of H$_2$ emission connects knot No.\ 41
to MMS~2/3 and it is possible that this structure is a faint H$_2$
jet that is about $15\arcsec $ north of axis $D$.  The orientation of
the bright bow shock (42) and a group of fainter knots around it (the
bow-shaped no.\ 50, and possibly 36) lies $20\arcsec $ south of axis
$D$.  The only candidate source of this flow at present is a 2~$\mu$m
binary that lies on this axis at ($\alpha , \delta$) = [$\rm { 05^h
32^m 55.^s4, -5^{\circ} 02\arcmin 34\arcsec }$].

A network of faint diffuse filaments at P.A.\ $\sim
45^{\circ}$--$60^{\circ}$ form a fan-shaped structure northeast of
OMC-3 (MMS~1 through 6).  Without spectroscopy of the individual
streamers, it is not possible to determine whether they are excited
by shocks or by UV-induced fluorescence from an external ionizing
source, possibly embedded in the HII region NGC~1977 to the north.

\subsection{Vicinity of Haro 5a/6a}
The double-lobed bipolar reflection nebula centered on
IRAS~05329-0505 (MMS~7; Chini et al.\ 1997) in the southern part of
OMC-3 is Haro 5a/6a (Haro 1953).  Wolstencroft et al.\ present
imaging polarimetry that demonstrates that the {\it IRAS} source
illuminates the reflection nebulae and attribute non-centrosymmetric
polarization to a tilt in the obscuring circumstellar torus between
the two reflection lobes.  The western lobe of Wolstencroft's nebula
contains a knot of H$_2$ emission (No.\ 27) which coincides with a
collimated Herbig-Haro flow (HH~294 [RBD97]; flow $F$ in Figure~4)
emerging along the axis of the nebula on deep optical images in
H$\alpha$ and [S~II] (RBD97).  In the continuum subtracted image this
knot is the brightest member of a chain that extends back towards the
core of the nebula.

A source not seen in our images, nor in Chini et al.'s molecular maps
drives flow $G$, which includes knot 75 and the large bow
shock-shaped knot 80.  The flow axis intersects with HH~42 and HH~128
(Ogura \& Walsh 1991) 5\arcmin\ to the east, and the faint complex of
HH~295 at a comparable distance from the molecular ridge to the west
(RBD97).  A second E-W large-scale flow is suggested by HH~41 lying
2\arcmin\ to the north of HH~42.  The H$_2$ counterparts to these HH
objects have been found (Yu et al.\ 1997), which when linked
together, outline parsec scale infrared outflows emanating from
OMC-2/3.

The source MMS~8, located $\sim 1\arcmin $ south of
IRAS~05329-0505 powers a spectacular east-to-west molecular outflow
seen in a $J = 2-1$ $^{12}$CO map presented by Chini et al.\ (1997).
The western (blueshifted) lobe of this outflow coincides with a
region completely devoid of stars or diffuse emission in our images.
However, the eastern (redshifted lobe) terminates near an arc of
H$_2$ emission (Nos.\ 78, 79) and knot 75 lies near the northern
boundary of the eastern CO outflow lobe.  The CO map of Chini et al.\
shows a second bipolar CO outflow in this region centered on the
source MMS~10, oriented NE-SW, with a redshifted lobe that coincides
with the redshifted lobe of the MMS~8 CO flow.  Moreover, knots 78
and 79 comprise a distinct bow shock with an axis of symmetry that
indicates excitation by an outflow source located to the southwest.
This structure is more likely to be associated with flows $I$
or $J$ (see below) than flow $F$ or the MMS~10 CO flow.

A collimated E-W chain of H$_2$ knots (flow $H$) terminates 3\arcmin\
west of MMS~9 in a bright bow shock at $\delta$ = $-05^{\circ}
08\arcmin $ (No.\ 5).  A second bright bow (No.\ 4) and the faintly
bow-shaped knot 39 may delineate a second flow nearly parallel to but
about 10\arcsec\ south although precession or fragmentation of a
single flow cannot be ruled out.  Knots 4 and 5 coincide with the
bright objects HH~357 in the optical (R97).  The $H$ flow is
displaced south of the blue-shifted lobe of the bipolar CO outflow
powered by MMS~8  by $\sim 1\arcmin $.  No counterpart to the bright
bow shocks are seen east of this source although a faint knot
(No.\ 70) and the HH object HH~287 (R97) which coincides with it lie
in this general region.

Flow $E$ consists of a diagonal chain of emission knots spaced at
regular intervals along a curvilinear path (knots 21, 24, 31, 33, 38,
59, and possibly 76) at P.A.\ $\sim ~40^{\circ}$.  This string passes
close to MMS~8 and 9, and its south end lies to the southwest of
FIR~1c which coincides with a faint reflection nebula between knots
31 and 24.  If the faint 2~$\mu$m source which coincides with FIR~1c
powers the outflow, then knots 24 and 21 likely trace a counterflow.
Knots 21 and 24 coincide with the HH~385, providing evidence that the
portion of the flow lying to the southwest of the 2~$\mu$m source is
less obscured.

\subsection{OMC-2.}  The FIR~1a, b, and c, FIR~2, and IRAS~05329-0508
complex contains a panoply of knots and streamers which most likely
trace several overlapping flows.  A bright 2~$\mu$m continuum source,
lying about 20\arcsec\ east of FIR~1a, is centered on a network of
H$_2$ streamers oriented at P.A.\ = 50$^{\circ}$ that resembles a
hollow tube of emission.  This is flow $I$ which consists of knots 6,
23, 49, 53, 58, 74, and 77.  Knot 58, which lies close to the
continuum source, coincides with HH~383. At faint levels, H$_2$
emission is nearly continuous from knot 74 towards the southwest to
knot 6 for a total length of over 4\arcmin .

The OMC-2 region contains the spectacular and well-defined outflow
$J$ that contains bright, distinct, multiple bow shocks (67, 68, 69,
71, 72, 73).  This flow can be traced back to the IRS~4N and IRS~4S
complex (Pendleton et al.\ 1986) which is coincident with FIR~3
(Mezger, Wink, \& Zylka 1990).  A molecular outflow that runs mainly
N-S but with a blue-shifted component at roughly ${\rm P.A.} =
30^{\circ}$ is centered on IRS~4N/4S in the $^{12}$CO maps by Fischer
et al.\ (1985).  A weak 22~GHz H$_2$O maser is also associated with
IRS~4N (Genzel \& Downes 1979).  The axis of flow $J$ passes near the
bow shock consisting of knots 78 and 79 in the north.  A small bright
bow shock-shaped knot (No.\ 14) lies on this axis in the south at the
same projected distance from FIR~3.  Knots 18 and 20 lie close to the
axis of $J$ between knot 14 and FIR~3.  If these are all associated
with a single flow, then $J$ is nearly 10\arcmin\ (1.3 pc) long.

OMC-2 also contains the bright compact E-W H$_2$ jet, $K$, which
consists of a collimated streamer, No.\ 52 and possibly knot 51 west
of IRS~2 (Rayner et al.\ 1989).  This jet is also visible in the red
lines of [S~II] and is designated HH~384 (R97).  There are a number
of knots (Nos.\ 23, 37, 45) in the region between IRS~2 and FIR~1
that are difficult to associate with any particular source or flow
due to the high degree of complexity and confusion.  Knot 45
coincides with FIR~2, and 51, 45, and 34 in the north form a linear
chain, $N$.  Another possible flow from FIR~2 is flow $O$ (Nos.\ 45,
37, and 23).

Complex filamentary H$_2$ emission south of OMC-2 may in part trace
fluorescent H$_2$ emission excited by UV photons from the H~II region
M43.  However, some of these streamers lie parallel to flow $J$.
Some bright knots in this region (such as knots 62, 66 and 14)
resemble compact bow shocks which may trace additional flows.  A
linear chain of such knots (9, 10, 19, 22, 26, and 35) forms flow $L$
which appears to radiate away from FIR~6b and another (62, 60, 64,
and 66) forms flow $M$.

\subsection{H$_2$ Luminosity.} For the best defined outflows, the
total 2.12~$\mu$m $v = 1$--$0 ~S(1)$ emission line luminosity is
$L_{2.12} \sim 0.2 L_{\odot}$.  For a H$_2$ vibrational excitation
temperature of $T_e \sim 2000$~K, the total luminosity for H$_2$
emission is 10 times that of the $S(1)$ line (e.g., Scoville et
al.\ 1982).  Assuming a 2.2~$\mu$m extinction of $A_{2.2} = 1$, the
extinction-corrected H$_2$ luminosity is $L_{\hbox{\rm H$_2$}} \sim 5
L_{\odot}$.  If the flux contribution from fluorescence is minimal,
this is a strong lower limit on the rate at which mechanical energy
is injected into the molecular cloud since H$_2$ is but one of many
important coolants that radiate away thermal energy in shocks.

\section{Outflows and Star Formation in the OMC-2/3 Region}

The Orion Nebula has produced on the order of 500 to 1,000 young
stars, including a number of high mass ones in the last $10^6$
years.  Although much less luminous, the OMC-2 and 3 cores and the
dense ridge of molecular gas that extends north of the Orion Nebula
appears to be an extremely active site of on-going star formation
that contains over a dozen class-0 protostars, dozens of active
outflows, and perhaps hundreds of more evolved young stars that have
formed within the last few million years.  The large number of active
flows and sub-mm sources indicates that this region is continuing to
form stars at a high rate.  Furthermore, the level of outflow
activity indicates that this region is undergoing a `microburst' of
star formation that is likely to forge a gravitationally unbound
cluster richer than NGC~1333 and IC348 in Perseus.  The large number
of deeply-embedded class-0 sources indicates that this portion of
Orion is less evolved than either NGC~1977, the Orion Nebula, or the
NGC~1333 region in the Perseus cloud where dozens of Herbig-Haro
flows are visible (Bally, Devine, \& Reipurth 1996).

Assuming that the phase during which an outflow is observable only by
its H$_2$ or CO emission (as opposed to shocks visible as Herbig-Haro
objects) lasts $\sim 3 \times 10^4$ years, comparable to the duration
of class-0 YSOs, then the formation rate of both H$_2$ flows and
class-0 sources must be about 30 to 40 per $10^5$ years.  This rate
of star formation can produce all of the 300 to 400 stars observed in
2~$\mu$m images of this region (our images contain about 200) in
about $10^6$ years.  For every $n = 10$ flows, with each flow
sweeping out an area $A_{-2}$ scaled to $10^{-2}$~pc$^2$, and with
velocities $v_{30}$ scaled to a typical flow speed of
$30$~km~s$^{-1}$, the timescale for reprocessing the local molecular
cloud material is $10^5 n_{10}^{-1} A_{-2}^{-1}
v_{30}^{-1}$~yr~pc$^{-3}$.  The cumulative effect of jets and
outflows from sustained star formation must be an important source of
dissociating shocks, and turbulent motions, and must play a crucial
role in the dynamics, chemistry, and evolution of star formation
within the cloud.

Near-IR H$_2$ emission is a potent tracer of outflow activity,
especially in regions where dozens to hundreds of stars are forming
per cubic parsec.  CO searches for outflows are hampered by large
beam sizes and the resulting high degree of source confusion that
prevents the recognition of individual, overlapping outflows.
Extremely young stars are often so obscured that even mature outflows
remain invisible at optical wavelengths.  Although Orion is one of
the most studied sites of star formation, both optical searches for
shock excited Herbig-Haro emission, and bipolar CO outflows have
utterly failed to reveal the splendor of star formation activity that
is now igniting inside the northern portion of the Orion~A cloud.  

\acknowledgments
This work was supported by NASA grant NAGW 3192 (LTSA).  The authors
would like to thank B.\ Ali for providing a complete source list
from his 1995 survey, and B.\ Reipurth for helpful comments.
\clearpage




\newpage

\small

\begin{deluxetable}{lrrcr}
\tablecolumns{4}
\tablewidth{0pc}
\tablecaption{OMC-2/OMC-3 H$_2$ Outflows}
\tablehead{ No.\tablenotemark{a} &$\alpha$(1950)\tablenotemark{b} &$\delta$(1950)\tablenotemark{b} & Outflow\tablenotemark{c} \\
 & ( $^h$~$^m$~$^s$ ) & ($^{\circ}$ \arcmin\  \arcsec\ ) & }
\startdata

1  &  5 32 40.4 & -5 02 42 & B \nl
2  &  5 32 44.1 & -5 02 30 & B \nl
3  &  5 32 45.9 & -5 02 29 & B \nl
4\tablenotemark{d}  &  5 32 47.9 & -5 08 07 & H \nl
5\tablenotemark{d}  &  5 32 48.7 & -5 07 56 & H \nl
6  &  5 32 48.8 & -5 11 45 & I \nl
7  &  5 32 49.2 & -5 02 29 & B \nl
8\tablenotemark{d}  &  5 32 50.5 & -5 07 45 & H \nl
9  &  5 32 50.6 & -5 14 40 & L \nl
10 &  5 32 51.0 & -5 14 35 & L \nl
11 &  5 32 51.5 & -5 13 35 &   \nl
12 &  5 32 51.7 & -5 13 43 &   \nl
13 &  5 32 52.3 & -5 03 12 & C \nl
14\tablenotemark{d} &  5 32 52.6 & -5 16 05 & J \nl
15\tablenotemark{d} &  5 32 53.1 & -5 07 41 & H \nl
16\tablenotemark{d} &  5 32 53.7 & -5 07 51 & H \nl
17 &  5 32 53.8 & -5 03 12 & C \nl
18 &  5 32 54.1 & -5 14 50 & J \nl
19 &  5 32 54.3 & -5 14 07 & L \nl
20 &  5 32 54.6 & -5 14 30 & J \nl
21 &  5 32 54.6 & -5 09 21 & E \nl
22 &  5 32 54.9 & -5 13 59 & L \nl
23 &  5 32 55.0 & -5 10 33 & O \nl
24 &  5 32 55.0 & -5 09 13 & E \nl
25\tablenotemark{d} &  5 32 55.1 & -5 07 44 & H \nl
26 &  5 32 55.1 & -5 14 04 & L \nl
27 &  5 32 55.3 & -5 05 56 & F \nl
28 &  5 32 55.4 & -5 03 10 & A \nl
29 &  5 32 55.6 & -5 07 46 & H \nl
30 &  5 32 55.6 & -5 03 25 & A \nl
31 &  5 32 55.7 & -5 08 57 & A \nl
32 &  5 32 55.7 & -5 03 39 & E \nl
33 &  5 32 55.7 & -5 08 51 & E \nl
34 &  5 32 55.7 & -5 09 46 & N \nl
35 &  5 32 55.9 & -5 13 53 & L \nl
36 &  5 32 55.9 & -5 02 15 & D \nl
37 &  5 32 55.9 & -5 10 37 & O \nl
38 &  5 32 56.0 & -5 08 37 & E \nl
39 &  5 32 56.1 & -5 07 56 & H \nl
40 &  5 32 56.1 & -5 07 48 & H \nl
41 &  5 32 56.5 & -5 02 00 & D \nl
42 &  5 32 56.7 & -5 02 19 & D \nl
43 &  5 32 56.7 & -5 07 43 & H \nl
44 &  5 32 56.7 & -5 12 36 &   \nl
45 &  5 32 56.7 & -5 10 33 & N \nl
46 &  5 32 56.9 & -5 05 56 & F \nl
47 &  5 32 57.0 & -5 15 28 &   \nl
48 &  5 32 57.2 & -5 07 47 & H \nl
49 &  5 32 57.3 & -5 09 55 & I \nl
50\tablenotemark{d} &  5 32 57.4 & -5 02 09 & D \nl
51 &  5 32 57.5 & -5 11 10 & N \nl
52 &  5 32 57.7 & -5 11 16 & K \nl
53 &  5 32 57.8 & -5 10 02 & I \nl
54 &  5 32 58.0 & -5 07 47 & H \nl
55 &  5 32 58.1 & -5 16 24 &   \nl
56 &  5 32 58.1 & -5 16 18 &   \nl
57 &  5 32 58.3 & -5 16 14 &   \nl
58 &  5 32 58.7 & -5 09 45 & I \nl
59 &  5 32 59.5 & -5 07 11 & E \nl
60 &  5 32 59.5 & -5 14 12 & M \nl
61 &  5 32 59.8 & -5 09 05 &   \nl
62 &  5 32 59.8 & -5 13 43 & M \nl
63 &  5 32 59.9 & -5 08 55 &   \nl
64 &  5 32 59.9 & -5 14 33 & M \nl
65 &  5 33 00.0 & -5 11 17 & J or K \nl
66 &  5 33 00.0 & -5 14 40 & M \nl
67\tablenotemark{d} &  5 33 00.6 & -5 10 56 & J \nl
68\tablenotemark{d} &  5 33 01.2 & -5 10 55 & J \nl
69\tablenotemark{d} &  5 33 01.4 & -5 10 35 & J \nl
70 &  5 33 02.0 & -5 07 40 & H & \nl
71\tablenotemark{d} &  5 33 02.2 & -5 10 51 & J \nl
72\tablenotemark{d} &  5 33 02.5 & -5 10 15 & J \nl
73\tablenotemark{d} &  5 33 02.5 & -5 10 27 & J \nl
74 &  5 33 03.8 & -5 08 16 & I & \nl
75 &  5 33 03.8 & -5 06 34 & G & \nl
76 &  5 33 04.1 & -5 05 08 & E & \nl
77\tablenotemark{d} &  5 33 04.5 & -5 08 13 & I \nl
78\tablenotemark{d} &  5 33 06.6 & -5 07 06 & J \nl
79\tablenotemark{d} &  5 33 08.9 & -5 06 58 & J \nl
80\tablenotemark{d} &  5 33 14.2 & -5 06 34 & G \nl
\enddata

\tablenotetext{a}{We suggest that authors refer to our H$_2$ knots
     using the knot numbers given in the first column plus the prefix
     {\it YBD} (for Yu, Bally, \& Devine); e.g., {\it YBD-1}.}

\tablenotetext{b} {The coordinates are accurate to about 1.0\arcsec .}

\tablenotetext{c} {The candidate exciting sources for the outflows
     are: $B$, source associated with MMS~2/MMS~3; $C$, source
     associated with MMS~5; $E$, FIR~1c; $F$, IRAS~05329-0505/MMS~7;
     $G$, source not visible in NIR nor FIR, and not associated with
     a mm core; $H$, source associated with MMS~9 or MMS~10?; $I$,
     source associated with IRS~1a; $J$, IRS~4N/FIR~3; $K$, IRS~2;
     $L$, FIR~6b; $N$, FIR~2?; $O$, FIR~2?}

\tablenotetext{d} {Knot is bow shock-shaped or part of a larger
     bow shock-shaped structure.}

\end{deluxetable}

\newpage
\section*{FIGURE CAPTIONS}

\bigskip {\bf Figure 1: (Plate 000)}
A $6\arcmin \times 16\arcmin $ KPNO 2.1 meter image of the
OMC-2/OMC-3 region showing the emission in a 1\% passband centered on
2.14~$\mu$m which contains no bright emission lines.  {\it Diamonds}
mark the location of 1.3~mm cores (Chini et al.\ 1997); {\it
triangles} mark the locations of far-infrared cores (Mezger, Wink, \&
Zylka 1990); {\it circles} mark the locations of {\it IRAS} sources;
{\it plus signs} mark the locations of OMC-2 infrared sources (Gatley
et al.\ 1974; Pendleton et al.\ 1986).

\bigskip {\bf Figure 2: (Plate 000)}
A $6\arcmin \times 16\arcmin $ KPNO 2.1 meter image of the
OMC-2/OMC-3 region showing the emission in a 1\% passband centered on
the 2.122~$\mu$m $v=1$--$0 ~S(1)$ line of H$_2$.  Symbols are for
sources identified in Figure~1.

\bigskip {\bf Figure 3: (Plate 000)}
A difference image of the OMC-2/OMC-3 region showing the 2.122~$\mu$m
$v=1$--$0 ~S(1)$ emission after a 2.14~$\mu$m image, normalized to the
bright stars, has been subtracted.  The line emission knots show up
lighter in color.  Symbols are for sources identified in Figure~1.

\bigskip {\bf Figure 4: (Plate 000)}
A 2.122~$\mu$m $v=1$--$0 ~S(1)$ line image of H$_2$ showing details
of the OMC-2/OMC-3 field.  Emission line bright knots listed in
Table~1 are shown with the notation given in column 1 of Table~1.
These knots do not have counterparts in the 2.14~$\mu$m continuum
image, or appear prominently in the continuum-subtracted image (see
Figure 3).  The positions of suspected outflows are also plotted as
dashed lines.  Symbols are for sources identified in Figure 1.



\begin{references}


\refpar {Ali, B.\ \& Depoy, D.\ L. 1995, \aj, 109, 709}

\refpar {Bally, J., Devine, D., \& Reipurth, B. 1996, \apjl, 473, L49}

\refpar {Bally, J.\, Devine, D., Hereld, M., \& Rauscher, B.\ J.
  1993, \apj, 418, 75}

\refpar {Bally, J., Langer, W.\ D., Stark, A.\ A., \& Wilson, R.\ W.
  1987, \apjl, 312, L45}

\refpar {Casali, M.\ M.\ \& Hawarden, T.\ G., JCMT-UKIRT Newsletter, No.\ 3,
  Aug.\ 1992, 33}


\refpar {Cesaroni, R., \& Wilson, T.\ L. 1994, \aap, 281, 209}

\refpar {Chini, R., Reipurth, B., Ward-Thompson, D., Bally, J.,
  Nyman, L.-\AA., Sievers, A., \& Billawala, Y. 1997, \apjl, 474, L135}

\refpar {Cohen, J.\ G., \& Frogel, J.\ A.  1977, \apj, 211, 178}

\refpar {Cooper, D.\ E., Bui, D.\ Q., Bailey, R.\ B., Kozlowski, L.\ J., \&
  Vural, K. 1993, Proc. SPIE, 1946, 170}

\refpar {Davis, C.\ J.\ \& Eisl\"offel, J.  1995, \aap, 300, 851}

\refpar {Dutrey, A.\, Duvert, G., Castets, A., Langer, W.\ D., Bally, J.,
  \& Wilson, R.\ W. 1993, \aap, 270, 468}

\refpar {Fischer, J., Sanders, D.\ B., Simon, M., \& Solomon, P.\ M. 1985,
  \apj, 293, 508}

\refpar {Gatley, I.\, Becklin, E.\ E., Matthews, K., Neugebauer, G.,
  Penston, M.\ V., \& Scoville, N. 1974, \apjl, 191, L121}

\refpar {Genzel, R.\ \& Downes, D. 1979, \aap, 72, 234}


\refpar {Hodapp, K.-W. \& Ladd, E.\ F. 1995, \apj, 453, 715}

\refpar {Johnson, J.\ J.\, Gehrz, R.\ D., Jones, T.\ J., Hackwell, J.\ A.,
  \& Grasdalen, G.\ L. 1990, \aj, 100, 518}

\refpar {Jones, T.\ J., Mergen, J., Odewahn, S., Gehrz, R.\ D., Gatley, I.,
  Merrill, K.\ M., Probst, R., \& Woodward, C.\ E. 1994, \aj, 107, 2120}

\refpar {McCaughrean, M.\ J., Rayner, J.\ T., \& Zinnecker, H. 1994, \apj,
  436, L189}

\refpar {Mezger, P.\ G., Wink, J.\ E., \& Zylka, R. 1990, \aap, 228, 95}

\refpar {Ogura, K.\ \& Walsh, J.\ R. 1991, \aj, 101, 185}

\refpar {Pendleton, Y., Werner, M.\ W., Capps, R., \& Lester, D. 1986, \apj,
  311, 360}

\refpar {Rayner, J., McLean, I., McCaughrean, M., \& Aspin, C. 1989,
  \mnras, 241, 469}

\refpar {Reipurth, B., Bally, J., \& Devine, D. 1997, \aj, in press (RBD97)}

\refpar {Reipurth, B. 1997, A General Catalogue of Herbig-Haro Objects,
  electronically published via anonymous ftp to ftp.hq.eso.org,
  directory /pub/Catalogs/Herbig-Haro (R97)}


\refpar {Rieke, M.\ J., Rieke, G.\ H., Green, E.\ M., Montgomery, E.\ F.,
  \& Thompson, C.\ L. 1993, Proc. SPIE, 1946, 179}

\refpar {Scoville, N.\ Z., Hall, D.\ N.\ B., Kleinmann, S.\ G., \&
  Ridgway, S.\ T.  1982, \apj, 253, 136}

\refpar {Tatematsu, K. et al., 1993, \apj, 404, 643}

\refpar {Thronson, H.\ A., Harper, D.\ A., Keene, J., Loewenstein, R.\ F.,
  Moseley, H., \& Telesco, C.\ M. 1978, \aj, 83, 492}

\refpar {Thronson, H.\ A., \& Thompson, R.\ I. 1982, \apj, 254, 543}


\refpar {Wolstencroft, R.\ D., Scarrott, S.\ M., Warren-Smith, R.\ F.,
  Walker, H.\ J., Reipurth, B., \& Savage, A. 1986, \mnras, 218, 1P}

\refpar {Yu, K.\ C., Reipurth, B., Billawala, Y., Bally, J., \& Devine, D.
  1997, in preparation}

\refpar {Zinnecker, H., McCaughrean, M.\ J., \& Rayner, J.\ T., 1997, in
  Low Mass Star Formation from Infall to Outflow, Poster Proceedings of
  the IAU Symp.\ No.\ 182, 20--24 Jan.\ 1997, eds.\ F.\ Malbet and
  A.\ Castets, 198}

\end{references}
\end{document}